\documentclass{article}
\usepackage{spconf,amsmath,graphicx}
\usepackage{here}
\usepackage{amsmath}

\usepackage{amsmath,bm,amssymb,url,cite}
\usepackage{hyperref}
\usepackage{multirow}
\def\Underline{\setbox0\hbox\bgroup\let\\\endUnderline}
\def\endUnderline{\vphantom{y}\egroup\smash{\underline{\box0}}\\}
\def\|{\verb|}

\def\R{\mathbb{R}}
\def\psil{p^{(\text{sil})}}

\def\thline{\noalign{\hrule height 1.0pt}}
\def\tthline{\noalign{\hrule height 1.4pt}}
\title{DIFFERENTIABLE digital signal processing MIXTURE MODEL for SYNTHESIS PARAMETER EXTRACTION from Mixture of Harmonic Sounds}

\name{\begin{tabular}{c}Masaya Kawamura$^1$, Tomohiko Nakamura$^1$, Daichi Kitamura$^2$,  \\ Hiroshi Saruwatari$^1$, Yu Takahashi$^3$, Kazunobu Kondo$^3$\end{tabular}}
\address{$^1$ The University of Tokyo, Tokyo, Japan \\ $^2$ National Institute of Technology, Kagawa College, Kagawa, Japan \\ $^3$ Yamaha Corporation, Shizuoka, Japan \\
}

\begin{document}
\ninept

\maketitle

\begin{abstract}
A differentiable digital signal processing (DDSP) autoencoder is a musical sound synthesizer that combines a deep neural network (DNN) and spectral modeling synthesis. It allows us to flexibly edit sounds by changing the fundamental frequency, timbre feature, and loudness (synthesis parameters) extracted from an input sound. However, it is designed for a monophonic harmonic sound and cannot handle mixtures of harmonic sounds. In this paper, we propose a model (DDSP mixture model) that represents a mixture as the sum of the outputs of multiple pretrained DDSP autoencoders. By fitting the output of the proposed model to the observed mixture, we can directly estimate the synthesis parameters of each source. Through synthesis parameter extraction experiments, we show that the proposed method has high and stable performance compared with a straightforward method that applies the DDSP autoencoder to the signals separated by an audio source separation method.

\end{abstract}


\section{Introduction}
\label{sec:intro}

Musical instrument sound synthesizers based on deep neural networks (DNNs) have been actively studied \cite{donahue2018adversarial,newt,htp,drumgan,u_net, crash}.
Such synthesizers can generate high-quality musical instrument sounds and also allow us to edit the sounds by appropriately changing their inputs and parameters.
Recently, an approach called differentiable digital signal processing (DDSP) has gathered attention \cite{ddsp_ref}.
This approach utilizes classical signal processing components for a DNN-based sound synthesizer and enables us to train the synthesizer in an end-to-end manner.
The DDSP autoencoder is one of the state-of-the-art DNN-based synthesizers categorized in this approach \cite{ddsp_ref}.
It reconstructs an input audio signal by a classical signal processing technique called spectral modeling synthesis (SMS) \cite{sms}, which separately models harmonic and inharmonic parts of the signal.
The control signals of the SMS are computed by a DNN.
As a latent representation, this DNN transforms the input into three interpretable parameters corresponding to pitch, timbre, and loudness: fundamental frequency ($F_0$), timbre feature, and loudness.
We call these parameters the synthesis parameters.
By appropriately changing the synthesis parameters, we can flexibly edit the pitch, timbre, and loudness of the input signal.

However, the DDSP autoencoder cannot be applied directly to a mixture of harmonic sounds because it is designed only for a monophonic harmonic signal.
One straightforward method to solve this problem is to separate the mixture into individual sources and apply the DDSP autoencoder to each of them.
Despite the recent progress of DNN-based audio source separation methods \cite{bunrireport}, 
it is difficult to always obtain separated signals indistinguishable from the clean ones.
Since the DDSP autoencoder is trained with clean musical instrument audio signals, the artifacts and interferer signals included in the separated signals can cause the performance degradation of the DDSP autoencoder.
In fact, the separated signals obtained with a state-of-the-art score-informed source separation method partly included the interferer signals, and the signals reconstructed by the DDSP autoencoder were considerably different from the target signals, which we will show later in section \ref{sec:eval}.
Furthermore, in practice, we often need to edit mixtures of sounds made by the same instruments. Although the separation of such mixtures has been studied recently \cite{A1,A2}, it is more difficult than the separation of sounds made by the different instruments.

In this paper, we propose a method for directly estimating the synthesis parameters of the individual sources from a mixture audio signal. We take not the \textit{separation-and-analysis} approach described above but an \textit{analysis-by-synthesis} approach.
That is, we construct a model that describes a generative process of the mixture, and we estimate the synthesis parameters by fitting the mixture generated with the model to an observed mixture.
By removing the synthesis parameter extraction part from the input in the DDSP autoencoder, we can use it to synthesize the source from the synthesis parameters.
The proposed model represents the mixture as the sum of the outputs of the source synthesizers driven with their own synthesis parameters. We call this model the \textit{DDSP mixture model}.
Using the pretrained source synthesizers, we fit the output of the proposed model to the observed mixture by a gradient descent algorithm.

Owing to the interpretability of the synthesis parameters,
we can use musical score information for the initialization of the synthesis parameters.
Recent source separation literature has shown that the use of score information improves the separation performance \cite{gakuhukouka1, gakuhukouka2, gakuhukouka3,simsd}, which may be true for our problem. We experimentally examine the effect of the score-based initialization of the synthesis parameters.
\section{RELATED WORKS}
\label{sec:relatedwork}
\subsection{DDSP Autoencoder}
\label{sec:ddsp}
\begin{figure}[t]
\centering 
\includegraphics[scale=0.29]{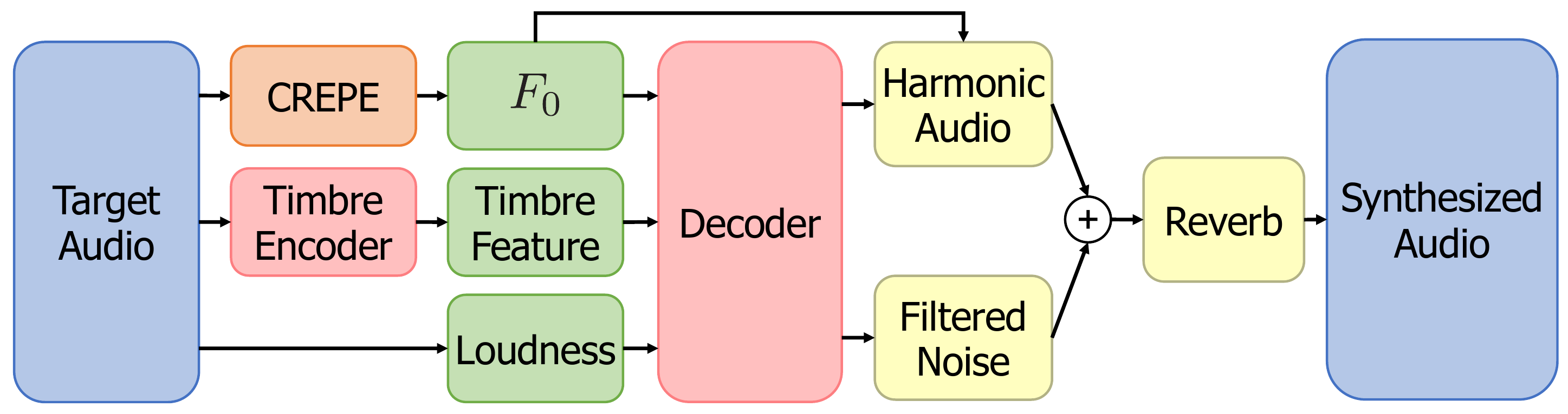}
\caption{Architecture of DDSP autoencoder. Red blocks consist of DNNs to be trained and “CREPE” block is pretrained DNN.}
\vspace{-9pt}
\label{fig:ddsp}
\end{figure}

The DDSP autoencoder consists of an encoder, a decoder, and an SMS module.
Fig. \ref{fig:ddsp} shows a schematic illustration of the architecture of the DDSP autoencoder.
The encoder extracts the synthesis parameters of $T$ frames from an input signal $\bm{x}\in \R^{N}$ with a length of $N$.
Let $t=1,\ldots,T$ be the frame index.
The $F_0$ at frame $t$, denoted by $f_t\geq 0$, is computed by a pretrained CREPE model \cite{crepe}, which is one of the state-of-the-art $F_0$ estimators. The timbre feature of size $D$, $\bm{z}_t\in\R^D$, is calculated by a timbre encoder, which computes mel-frequency cepstral coefficients (MFCCs) from the input signal and feeds them into a DNN.
The loudness $l_t\in\R$ is computed by applying A-weighting to the power spectrum of the input signal and taking its logarithm.
The decoder is a DNN that transforms the synthesis parameters into the control signals of the SMS module in frames. See \cite{ddsp_ref} for the detailed architecture of the decoder and timbre encoder.

The SMS module 
separately generates harmonic and inharmonic signals and adds them together.
The harmonic signal is generated as the sum of sinusoids with piecewise linear frequencies and amplitudes.
These frequencies are computed by linearly interpolating $f_{t}$ and its harmonics up to the signal time resolution. The amplitudes are the linearly interpolated versions of framewise amplitudes outputted by the decoder.
To generate the inharmonic signal, the decoder outputs the magnitude frequency responses of a time-varying finite impulse response filter in frames.
We apply a Hann window to the discrete Fourier transforms of these responses and convolve them with a white noise signal in the frequency domain.
A reverb module implemented by a convolutional layer is optionally applied to the sum of the harmonic and inharmonic signals and outputs a synthesized signal $\bm{\hat{x}}\in\R^{N}$.

The timbre encoder, decoder, and reverb module are trained so that the multiscale spectral loss between $\bm{x}$ and $\bm{\hat{x}}$ is minimized \cite{ddsp_ref}.
This loss uses short-time Fourier transforms (STFTs) of the two inputs with frames of $I$ different lengths.
It is defined as
\begin{align}
    L(\bm{x},\bm{\hat{x}}) &= \sum_{i=1}^{I}L_i(\bm{x},\bm{\hat{x}}), \label{eq:mss_loss}\\
    L_i(\bm{x},\bm{\hat{x}}) &= \lVert \mathcal{F}_i\bm{x} - \mathcal{F}_i\bm{\hat{x}}\rVert_1 + \lVert \log \mathcal{F}_i\bm{x} - \log \mathcal{F}_i \bm{\hat{x}}\rVert_1,
\end{align}
where $\mathcal{F}_i$ returns the magnitude STFT of the signal with the $i$th frame length.

\subsection{Audio Source Separation}
Most conventional music audio editing systems use audio source separation methods as preprocessors to extract the target sources from a mixture \cite{timbre_replacement, equalizer, changing_timbre, time_domain}. This approach can be applied to our problem.
The recent literature has shown that the use of musical scores enhances the separation performance \cite{gakuhukouka1,gakuhukouka2,gakuhukouka3,simsd}.
The method using nonnegative matrix factorization (NMF) presented in \cite{simsd} is one of the state-of-the-art score-informed source separation methods. This method trains NMF bases with isolated instrument sounds in advance and separates the input mixture while aligning the performance and score information.
Although the DNN-based methods show superior performance in the usual supervised source separation setting \cite{bunrireport}, in the score-informed setting, this NMF-based method works better than DNN-based methods \cite{simsd}.

The audio source separation method presented in \cite{gp} uses pretrained instrument sound synthesizers based on generative adversarial networks (GANs).
The GANs convert random vectors into audio signals.
These vectors are thus difficult to interpret and introduce prior musical knowledge into the inputs.
Furthermore, GANs are usually unstable during training, which requires painstaking hyperparameter exploration \cite{gan_overview}.
\section{PROPOSED METHOD}
\subsection{Motivation and Strategy}\label{sec:motivation}
\begin{figure*}[t]
\centering 
\includegraphics[scale=0.52]{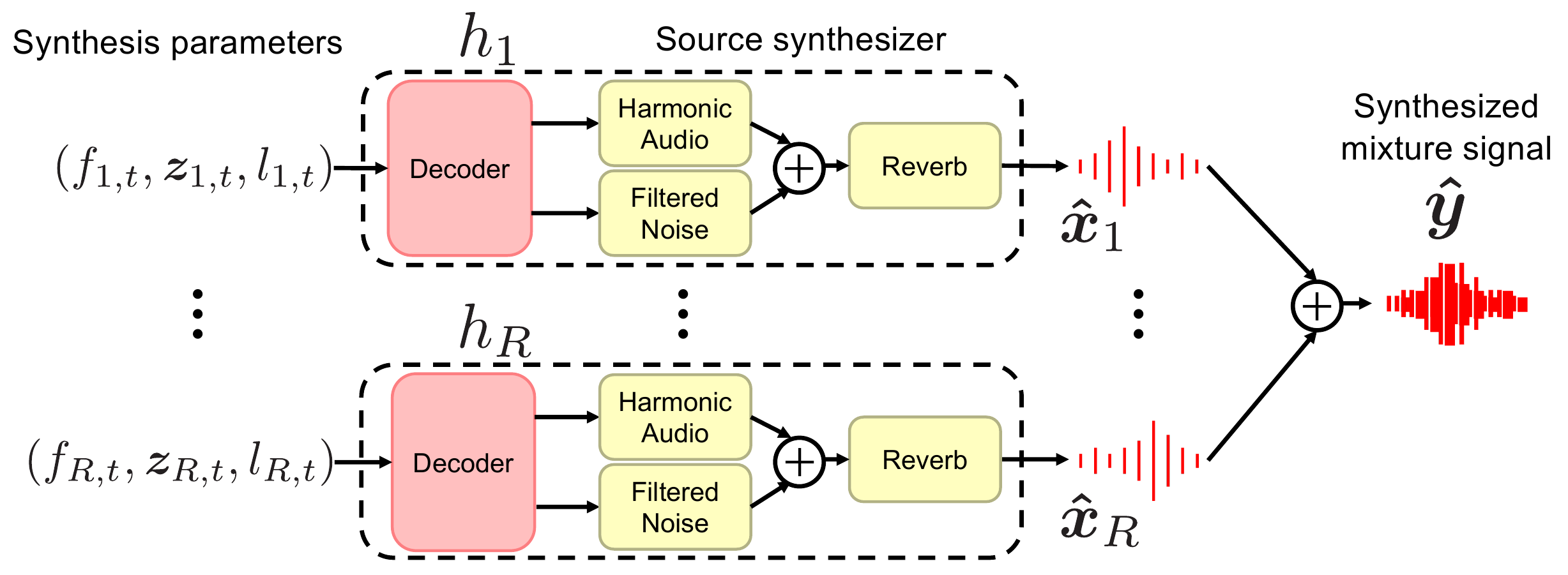}
\caption{DDSP mixture model with $R$ sources.}
\vspace{-9pt}
\label{proposed}
\end{figure*}
One straightforward approach to use the DDSP autoencoder for polyphonic audio signals is to decompose the mixture into the source signals and apply the DDSP autoencoder to them.
Since the DDSP autoencoder is trained with only clean instrument sounds, its synthesis performance is strongly affected by the artifacts and interferer signals included in the separated signals.
Although the introduction of DNNs has rapidly increased the performance of source separation methods \cite{bunrireport}, the separated signals obtained even with the latest methods often include artifacts and interferer signals, and sometimes lack part of the target source signals.
These separation failures lead to the performance degradation of the DDSP autoencoder, as we will show later in section \ref{sec:eval}.

To avoid this problem, we take an approach in which the synthesis parameters of the sources are directly extracted from the mixture.
We focus on the fact that the part subsequent to the encoder of the DDSP autoencoder can be seen as a source audio synthesizer using the synthesis parameters (see Fig.~\ref{fig:ddsp}). We call it the source synthesizer. Using multiple source synthesizers, we construct a generative model of the harmonic sound mixture as shown in Fig.~\ref{proposed}. We also formulate the synthesis parameter extraction problem as an inverse problem using the proposed model.
\subsection{DDSP Mixture Model}\label{sec:algo}
The proposed DDSP mixture model represents the mixture audio signal of $R$ harmonic sources as the sum of the outputs of the source synthesizers driven with source-specific synthesis parameters.
Let $r=1,\ldots,R$ denote the source index and $h_r$ represent the source synthesizer of source $r$.
To distinguish the synthesis parameters and synthesized signals of each source, we hereafter add a subscript $r$ to $f_t,\bm{z}_t,l_t,$ and $\bm{\hat{x}}$.
Fig. \ref{proposed} shows the architecture of the DDSP mixture model.
The synthesis parameters of source $r$, $\{f_{r,t},\bm{z}_{r,t},l_{r,t}\}_{t=1}^T$, are fed into $h_r$, and the synthesized signal of source $r$, $\bm{\hat{x}}_r$, is generated.
Adding all the synthesized source signals yields the synthesized mixture signal $\bm{\hat{y}}\in\R^{N}$:
\begin{align}
    \bm{\hat{y}} &= \sum_{r=1}^R\bm{\hat{x}}_r
    \label{eq:ddsp_mixture}, \\
    \bm{\hat{x}}_r &= h_r(\{f_{r,t},\bm{z}_{r,t},l_{r,t}\}_{t=1}^T).
    \label{eq:ddsp_r}
\end{align}
Note that although all $h_r$ are separately depicted in Fig. \ref{proposed}, we can use the same pretrained source synthesizer for all sources when the DDSP autoencoder is trained with multiple instrument sounds.

The DDSP mixture model describes the forward process of the generation of the harmonic sound mixture.
Thus, the synthesis parameter extraction problem amounts to the problem of finding the synthesis parameters of the sources so that they minimize the loss between the output of the DDSP mixture model $\bm{\hat{y}}$ and the observed mixture $\bm{y}\in\R^N$.
As a loss function, we can use the multiscale spectral loss defined in \eqref{eq:mss_loss}.
In summary, the problem of interest is formulated as
\begin{align}
\min_{\{f_{r,t},\bm{z}_{r,t},l_{r,t}\}_{r=1,t=1}^{R,T}} & L(\bm{y},\bm{\hat{y}}).
\label{eq:formulation}
\end{align}
Since $h_r$ and this loss are differentiable, we can use a gradient descent algorithm for this problem.
Note that that all $h_r$ are trained in advance and fixed during this minimization.
To distinguish the DDSP pretraining and this step, we call the latter the fitting step.
\subsection{Initialization of $F_0$ and Loudness Using Score Information}\label{sec:param_init}
Owing to the recent development of automatic music transcription \cite{B1,B2,B3}, accurate score information can be extracted from polyphonic music signals.
Since the proposed method uses the interpretable synthesis parameters, we can utilize the available score information for the initialization of $f_{r,t}$ and $l_{r,t}$.
For simplicity, the score information is given in a musical instrument digital interface (MIDI) format and is assumed to be aligned in time with the input mixture.
Let $p_{r,t}=-1,0,\ldots,127$ denote the MIDI note number of source $r$ at time $t$, where $p_{r,t}=-1$ means that there are no played notes at that time, i.e., silence.
By converting $p_{r,t}$ into the corresponding frequency, we can initialize $f_{r,t}$ as
\begin{equation}
    f_{r,t} = 
    \begin{cases}
    440\times 2^{(p_{r,t}-69)/12} & (p_{r,t}\geq0) \\
    440\times 2^{(\psil_r-69)/12} & (p_{r,t}=-1),
    \end{cases}
\end{equation}
where $p_r^{\text{(sil)}}$ is the time average of nonnegative $p_{r,t}$s.
The loudnesses $l_{r,t}$ are initialized with $l^{(\text{high})}$ for the active notes and $l^{(\text{low})}$ for the silences.
\begin{table*}[tb] 
\caption{Averages and standard errors of MAEs in $F_0$, MFCC, and loudness obtained with separation-based and proposed methods} 
\label{tab: result}
\hbox to\hsize{\hfil
\begin{tabular}{ccc|c|ccc}\tthline
Label & Instruments & Total dur. [s]& Method & $F_0$ [cent] & MFCC & Loudness $[$dB$]$\\ \hline
\multirow{3}{*}{Va./\rule{0.2cm}{0.15mm}.}
& & &SISS+DDSP  & $\mathbf{125\pm{35}}$  & $3.52\pm{0.12}$& $\mathbf{10.21\pm{1.75}}$\\
& Va./Db. (Mahler), Va./Fl. (Mahler) & 240 &SISS+Proposed & $129\pm{34}$ 
& $2.84\pm{0.13}$ & $12.06\pm{1.89}$ \\
& & &SI-Proposed  & $217\pm{48}$  & $\mathbf{2.79\pm{0.13}}$ & $10.86\pm{0.63}$\\\hline

\multirow{3}{*}{Fl./\rule{0.2cm}{0.15mm}.}
& & &SISS+DDSP  & $130\pm{45}$  & $2.04\pm{0.27}$& $33.43\pm{2.92}$\\
& Fl./Bn. (Mozart), Fl./Va. (Mahler) & 300 &SISS+Proposed & $132\pm{45}$ 
& $\mathbf{2.01\pm{0.18}}$ & $34.31\pm{3.28}$ \\
& & &SI-Proposed  & $\mathbf{90\pm{18}}$  & $2.25\pm{0.17}$ & $\mathbf{10.54\pm{0.59}}$\\\hline

\multirow{3}{*}{Db./\rule{0.2cm}{0.15mm}.}
& & &SISS+DDSP  & $1588\pm{1286}$  & $2.88\pm{0.15}$& $16.33\pm{1.96}$\\
& Db./Vc. (Beethoven), Db./Va. (Mahler) & 300 &SISS+Proposed & $902\pm{682}$ 
& $\mathbf{2.03\pm{0.15}}$ & $17.52\pm{2.61}$\\
& & &SI-Proposed  & $\mathbf{164\pm{56}}$  & $2.05\pm{0.15}$& $\mathbf{10.41\pm{0.61}}$\\\hline

\multirow{3}{*}{Vc./\rule{0.2cm}{0.15mm}.}
& & &SISS+DDSP  & $1572\pm{983}$  & $3.29\pm{0.09}$& $10.49\pm{2.04}$\\
& Vc./Db. (Beethoven), Vc./Bn. (Mozart) & 260 &SISS+Proposed & $1178\pm{777}$ 
& $2.32\pm{0.13}$ & $10.29\pm{2.29}$\\
& & &SI-Proposed  & $\mathbf{111\pm{24}}$  & $\mathbf{2.21\pm{0.11}}$ & $\mathbf{8.09\pm{0.26}}$\\\hline

\multirow{3}{*}{Bn./\rule{0.2cm}{0.15mm}.}
& & &SISS+DDSP  & $1043\pm{506}$  & $2.45\pm{0.11}$& $24.24\pm{1.65}$\\
& Bn./Fl. (Mozart), Bn./Vc. (Mozart) & 260 &SISS+Proposed & $911\pm{490}$ 
& $\mathbf{1.90\pm{0.13}}$ & $31.39\pm{1.94}$\\
& & &SI-Proposed  & $\mathbf{113\pm{10}}$  & $2.25\pm{0.15}$ & $\mathbf{10.47\pm{0.59}}$\\\thline

\multirow{3}{*}{Va./Va.}
& & &SISS+DDSP  & $\mathbf{140\pm{72}}$  & $3.83\pm{0.13}$& $13.38\pm{1.16}$\\
& Va. (Mahler)/Va. (Mozart) & 120 &SISS+Proposed & $146\pm{71}$ 
& $3.03\pm{0.12}$ & $13.16\pm{1.65}$ \\
& & &SI-Proposed  & $155\pm{38}$  & $\mathbf{2.89\pm{0.11}}$ & $\mathbf{11.84\pm{0.62}}$\\\hline

\multirow{3}{*}{Fl./Fl.}
& & &SISS+DDSP  & $129\pm{68}$  & $1.72\pm{0.27}$& $31.32\pm{4.05}$\\
& Fl. (Mahler)/Fl. (Mozart) & 120 &SISS+Proposed & $116\pm{57}$ 
& $1.62\pm{0.21}$ & $33.84\pm{4.33}$\\
& & &SI-Proposed  & $\mathbf{90\pm{26}}$  & $\mathbf{1.44\pm{0.23}}$ & $\mathbf{9.13\pm{0.52}}$\\\hline

\multirow{3}{*}{Db./Db.}
& & &SISS+DDSP  & $844\pm{754}$  & $2.33\pm{0.17}$ & $15.74\pm{2.86}$\\
& Db. (Beethoven)/Db. (Mahler) & 120 &SISS+Proposed & $877\pm{752}$ 
& $\mathbf{1.91\pm{0.13}}$ & $20.23\pm{3.32}$ \\
& & & SI-Proposed  & $\mathbf{195\pm{68}}$  & $1.97\pm{0.13}$ & $\mathbf{10.51\pm{0.81}}$\\\hline

\multirow{3}{*}{Vc./Vc.}
& & &SISS+DDSP  & $3438\pm{1851}$  & $2.79\pm{0.18}$ & $20.05\pm{3.21}$\\
& Vc. (Beethoven)/Vc. (Mahler) & 120 &SISS+Proposed & $2775\pm{1737}$ 
& $\mathbf{2.27\pm{0.14}}$ & $24.35\pm{3.85}$\\
& & &SI-Proposed  & $\mathbf{347\pm{144}}$  & $2.32\pm{0.16}$ & $\mathbf{12.06\pm{0.91}}$\\\hline

\multirow{3}{*}{Bn./Bn.}
& & &SISS+DDSP  & $1139\pm{527}$  & $3.36\pm{0.15}$ & $9.75\pm{2.02}$\\
& Bn./Bn. (Beethoven) & 180 &SISS+Proposed & $1151\pm{528}$ 
& $\mathbf{2.38\pm{0.15}}$ & $10.87\pm{2.16}$\\
& & &SI-Proposed  & $\mathbf{567\pm{248}}$  & $2.39\pm{0.16}$ & $\mathbf{8.54\pm{0.32}}$\\\tthline

\end{tabular}\hfil}
\vspace{-10pt}
\end{table*}

\section{EXPERIMENTAL EVALUATION}\label{sec:eval}
\subsection{Experimental Conditions}
To evaluate the effectiveness of the proposed method, we conducted synthesis parameter extraction experiments on mixtures of two harmonic sources.
We created test data using the PHENICX-Anechoic dataset \cite{gakuhukouka3,testdata2}.
This dataset includes separate audio recordings of musical instruments of four symphonies: Symphony no. 1, fourth movement by G. Mahler (Mahler), an aria of Donna Elvira from the opera Don Giovanni by W. A. Mozart (Mozart), and Symphony no. 7, first movement by L. van Beethoven (Beethoven).
It also includes time-aligned MIDI data.
The test data consisted of mixtures of two different instruments or two of the same instrument.
We used audio signals played with the viola (Va.), flute (Fl.), double bass (Db.), cello (Vc.), and bassoon (Bn.) and downsampled all audio signals to $16$ kHz.
We divided each mixture into $12$-s segments from the beginning to the end and applied synthesis parameter extraction methods to them.
The segments shorter than $12$ s were not used for the evaluation.

We compared the following three methods.

\noindent\textbf{SISS+DDSP:}
We separated the mixtures by the state-of-the-art score-informed source separation method (SISS) presented in \cite{simsd} and applied the DDSP autoencoder to the separated signals.
In our preliminary experiment, we found that this method provided higher separation performance than the recent DNN-based score-informed source separation method presented in \cite{si_dnn}.
We used the official implementation of SISS available at \url{https://github.com/AntonioJMM/OISS_Minus-One.github.io} and the same hyperparameters as those used in \cite{simsd}.

\noindent\textbf{SISS+Proposed:}
We initialized $f_{r,t}$ and $l_{r,t}$ with those obtained with SISS+DDSP and ran the proposed method.
The initial values of $z_{r,t}$ were drawn from a standard normal distribution.
We used the Adam optimizer and set its learning rate at $0.1$, which was decreased to $0.01$ at the $1000$th iteration and $0.001$ at the $2000$th iteration.
For the multiscale spectral loss, we used Hann windows of $8, 16, 32, 64, 128,$ and $256$-ms lengths and hop sizes of half of the corresponding frame lengths.

\noindent\textbf{SI-Proposed:}
This model is a score-informed (SI) version of the proposed method. We initialized $f_{r,t}$ and $l_{r,t}$ with the score information described in section \ref{sec:param_init} and ran the proposed method.
We experimentally determined that $l^{\text{(high)}}=-6$ and $l^{\text{(low)}}=-10$. The other conditions were the same as those of SISS+Proposed.

These methods used the same pretrained DDSP autoencoder.
It was trained using the University of Rochester multimodal music performance (URMP) dataset \cite{urmp}, which consists of $44$ classical chamber music pieces and audio signals played with $13$ musical instruments.
We used $35$ out of the $44$ music pieces as the training data (total $11523$ s) and divided all instrument signals into $12$-s segments.
The segments shorter than $12$ s were zero-padded up to $12$-s length.
We trained $3000$ epochs using the Adam optimizer with a learning rate of $0.001$. We used the multiscale spectral loss with the same frame lengths as in the fitting step.
The synthesis parameters were computed at $32$-ms intervals ($T=375$).

As evaluation measures, we used the mean absolute errors (MAEs) in $F_0$, MFCC, and loudness between the estimates and the ground truths extracted from the source signals.
The ground truths of $F_0$ were extracted using the CREPE model.
Following \cite{ddsp_ref}, the $F_0$ MAEs were computed at the frames for which the confidences of the ground truths of $F_0$ were greater than or equal to $0.85$.
Note that since the estimated $f_{r,t}$ values may be negative, we floored them with $10^{-7}$ Hz. The MFCC estimates were computed from the audio signals synthesized with the estimated synthesis parameters.
We calculated the MFCCs using the log-mel-spectrogram with $128$-ms frames, a $32$-ms hop size, and $128$ frequency bins ranging from $20$ to $8000$ Hz, and we used the first $30$ coefficients.

\subsection{Results}
Table~\ref{tab: result} shows the results of all methods, where the evaluation measures were computed at each segment and their averages and standard errors were computed.
Here, the instrument names followed by the slash and underline denote the results for the mixtures of the different instruments, and those followed by the slash and same instrument names denote the results for the mixtures of the same instruments.
The names inside the parentheses of the ``Instruments'' column indicate the music pieces in which the instrument audio signals are included.
Although SISS+DDSP provided moderate performance in terms of $F_0$ for the Va. and Fl. mixtures, it showed much lower performance for the other mixtures. A similar tendency was observed for MFCC and loudness.
When we listened to the separated signals of SISS+DDSP, we found that these signals lack part of the target sources and included artifacts and interferer signals.
We also found that some of the synthesized signals of SISS+DDSP were considerably different from the target source signals.
These results show that the separation-based method has unstable performance and that separation failures greatly degrade the synthesis parameter extraction performance.

Compared with SISS+DDSP, SISS+Proposed and SI-Proposed provided comparable and higher performances for most of the mixtures, particularly in terms of MFCC, showing the effectiveness of the proposed methods.
Although the SISS+Proposed performances of $F_0$ and loudness were still low for the mixtures that SISS failed to separate, SI-Proposed consistently provided higher performances for all measures.
Furthermore, we observed that the source signals synthesized by the proposed method had much more similar timbres to the target sources and audibly outperformed those obtained with SISS+DDSP.
These results clearly show that the proposed method works stably and effectively compared with the separation-based method. Some synthesized examples are available at \url{https://sarulab-audio.github.io/DDSP_Mixture_Model/}.

Importantly, SI-Proposed had much lower standard errors in $F_0$ and loudness than the other methods for most of the mixtures.
This result shows that the score information is useful for the proposed method and can decrease the deviations of the estimates.

\section{CONCLUSION}
We proposed the DDSP mixture model that represents the generation process of a mixture of harmonic audio signals, using part of the pretrained DDSP autoencoder as a source audio synthesizer.
We also developed a synthesis parameter extraction method by fitting the output of the DDSP mixture model to the observed mixture.
Through experiments using mixtures of sounds made by the different and same instruments, we showed that the proposed method outperforms a straightforward method that applies the DDSP autoencoder to signals separated with an existing audio source separation method.

\vfill\pagebreak
\bibliographystyle{IEEE}
\bibliography{strings,refs}

\end{document}